# Tunable Nanostructuring for van der Waals Materials

*Gleb I. Tselikov[1], Anton A. Minnekhanov[1], Georgy A. Ermolaev[1], Gleb V. Tikhonowski[1], Ivan S. Kazantsev[1], Dmitry V. Dyubo[1], Daria A. Panova[2], Daniil I. Tselikov[1], Anton A. Popov[1], Arslan B. Mazitov[1], Sergei Smirnov[1], Fedor Lipilin[3], Umer Ahsan[3], Nikita D. Orekhov[1], Ivan A. Kruglov[1], Alexander V. Syuy[1], Andrei V. Kabashin[4], Boris N. Chichkov[5], Zdenek Sofer[3], Aleksey V. Arsenin[1], Kostya Novoselov[6,7,8], and Valentyn S. Volkov[1]\**

[1]*Emerging Technologies Research Center, XPANCEO, Dubai Investment Park First, Dubai, United Arab Emirates*

[2]*Physics Department, King's College London, London WC2R 2LS, UK*

[3]*Department of Inorganic Chemistry, University of Chemistry and Technology Prague, Technická 5, 166 28 Prague 6, Czech Republic*

[4]*Aix-Marseille Université, CNRS, LP3, 13228 Marseille, France*

[5]*Institute of Quantum Optics, Leibniz Universität Hannover, 30167 Hannover, Germany*

[6]*The University of Manchester, National Graphene Institute, Oxford Rd, Manchester M13 9PL, UK*

[7] *Department of Materials Science and Engineering, National University of Singapore, Singapore, 03-09 EA, Singapore*

[8]*Chongqing 2D Materials Institute, Liangjiang New Area, Chongqing, 400714, China*

*\*Correspondence should be addressed to e-mail: vsv@xpanceo.com*

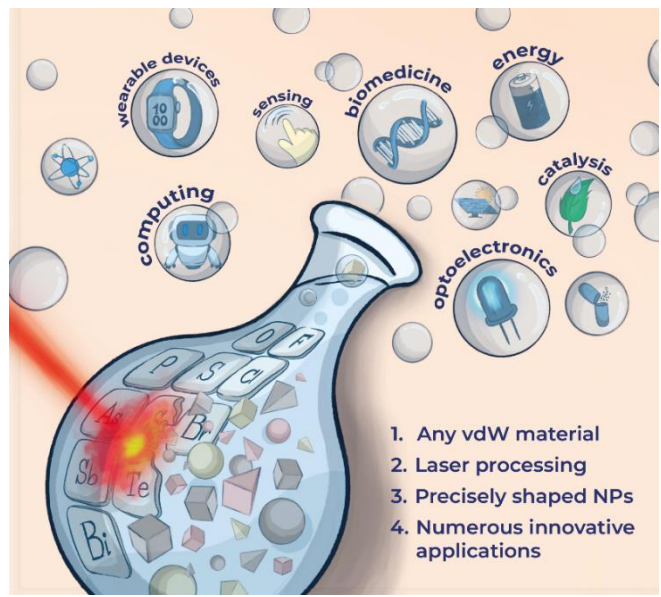

## Abstract

**Van der Waals (vdW) materials are becoming increasingly popular in scientific and industrial applications because of their unique mixture of record electronic, optical, and mechanical properties. However, nanostructuring of vdW materials is still in its infancy and strongly depends on the specific vdW crystal. As a result, the universal self-assembled**

technology of vdW materials nanostructuring opens vast technological prospects. **This work demonstrates an express and universal synthesis method of vdW nanoparticles with well-defined geometry using femtosecond laser ablation and fragmentation. The disarming simplicity of the technique allows us to create nanoparticles from over 50 vdW precursor materials covering transition metal chalcogenides, MXenes, and other vdW materials. Obtained nanoparticles manifest perfectly defined crystalline structures and diverse shapes, from nanospheres to nanocubes and nanotetrahedrons. Thus, our work provides a new paradigm for vdW nanostructuring with a vast potential of tunability for size, shape, and materials specific to the particular application.**

## Introduction

The era of 2D materials exploration began with the discovery of graphene, a prominent member of the van der Waals (vdW) family of materials. vdW materials have attracted significant attention in recent years due to their extraordinary optical, electronic, magnetic, and other properties [1–7]. These materials, characterized by weak interlayer vdW forces, facilitate the creation of atomically thin layers with remarkable characteristics [8,9] that can serve as building blocks for heterostructures with on-demand properties [10]. They have been reported to exhibit giant optical anisotropy with record high in-plane and low out-of-plane refractive indices [11–13]. This feature enables them to overcome the diffraction limit, rendering them superior for various nanophotonic applications, including photonic integrated circuitry [14,15], resonant nanoantennas [16], light guiding [17,18], and photodetectors [19]. Furthermore, this family encompasses a diverse range of electronic types, including record conductors such as MXenes [20], semiconductors such as two-dimensional chalcogenides and oxides [21], and insulators [22]. This versatility allows their use in direct ink printing to create 3D structures, which can be employed in constructing flexible and high-efficiency wireless electronics and energy storage devices [23,24].

However, the transition from fundamental research to practical applications has been hindered by several challenges. One major obstacle is the difficulty in nanostructuring these materials, which involves the synthesis of nanostructures with the desired size, shape, and properties and is essential for their integration into advanced technologies and devices. Implementing vdW materials-based devices is further complicated by substrate interactions, which can influence their properties and performance. Additionally, controlling the thickness of these materials during synthesis is a formidable task. Chemical vapor deposition (CVD), a common technique for growing thin films [25], often struggles to produce uniform and sufficiently thick vdW material films, which are crucial for specific applications. Other nanostructuring methods often necessitate bespoke protocols tailored to the unique properties of each material, requiring distinct conditions, such as temperature, pressure, and chemical environment, for the successful fabrication of nanostructures immobilized on a substrate [26–29]. Consequently, developing a universal and facile technique for nanostructuring vdW materials is highly desirable.

In this context, ultrashort pulsed laser synthesis of nanomaterials in liquids emerges as an attractive alternative to CVD methods. It stands out for its versatility and simplicity in producing nanoparticles (NPs) from a wide range of inorganic materials (metals, semiconductors, dielectrics) [30–35]. The absence of toxic precursors during the synthesis process and the use of high-quality materials enable the production of ultrapure, ligand-free NPs with controlled physicochemical properties [36,37]. NPs' size, morphology, chemical composition, surface properties, and optical characteristics can be precisely tuned by adjusting synthesis parameters such as liquid media type, pulse duration, radiation wavelength and energy, and focusing conditions [38–41]. The utilization of femtosecond (fs) pulses is particularly advantageous, as it preserves the original material's crystalline structure [30,42].

Furthermore, the pure radiation nature of laser-matter interaction and highly nonequilibrium nanocluster formation processes result in significant surface charge formed on the laser-synthesized NPs, ensuring their excellent colloidal stability [43]. Economically, the method is advantageous with advanced setups like flow cells, achieving productivity over 1 g/hour for popular NPs such as gold and ceramics [33,44], surpassing industrial wet chemistry methods, and making the technique easily scalable. Recent studies show that fs laser ablation can form spherical transition-metal dichalcogenide (TMDC) NPs, which retain the original crystalline structure and exhibit resonant optical properties [45]. Inspired by this result, we propose fs laser ablation in liquid as a versatile and effective tool to address the challenges associated with the nanostructuring of vdW materials. Its combination of simplicity, accessibility, high productivity, and the unique ability to precisely control the purity and properties of resulting nanomaterials renders this method highly valuable for a broad range of fields, including energy, catalysis, and sensing [46–49].

In this work, we report the development of a universal approach that enables a fast and efficient protocol for the synthesis of NPs with the desired size, morphology, optical characteristics, and large surface-to-volume ratio, which can potentially be allied for more than 5,000 members of the vdW family, overcoming the current limitations of material-specific protocols and facilitating the broader use of these materials in advanced technological applications. This approach allows the production of geometrically precise NPs from a broad range of 2D materials, covering a substantial part of the periodic table, which is schematically represented in Figure 1. Here, we present over 50 samples of geometrically precise NPs derived from various vdW precursor materials, including TMDCs, MXenes, and perovskites. We experimentally confirm that the structure of vdW NPs corresponds to the initial crystals and explore their extensive potential applications. The development and integration of these NPs into devices and systems holds significant promise for driving innovations and enhancing performance in energy, catalysis, medicine, electronics, and environmental remediation.

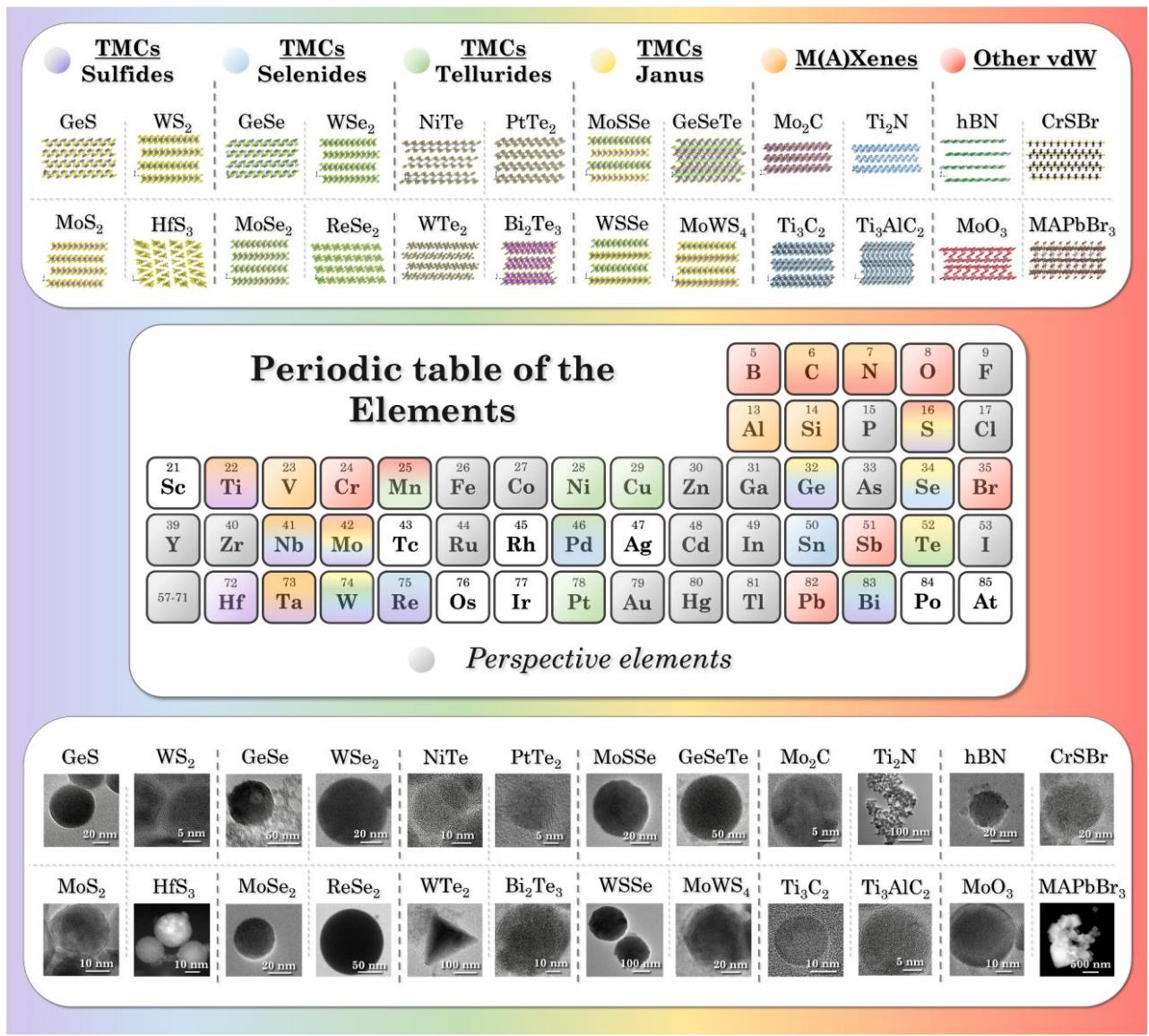

**Figure 1.** Schematic depiction of the periodic table (centered in a condensed format) with some prominent classes of layered materials that can be synthesized into regular NPs via laser ablation. The top part of the figure shows the crystalline structures of these materials, while the bottom part displays electron microscopy images of the NPs obtained by this method. Material classes are highlighted in color, and promising elements suitable for this technique are also marked.

## Experimental Results

### *Laser-driven synthesis of vdW NPs*

The synthesis of NPs through pulsed laser ablation and fragmentation in liquids marks a significant advancement in nanomaterials engineering. The use of liquid environments in laser processing offers novel benefits, such as reducing heat load, confining vapor and plasma, and producing ligand-free nanomaterials with controllable properties. Recently, this method has been demonstrated to be effective for synthesizing NPs from silicon, noble metals, and other materials [40,50–52]. Here, we elucidate the processes of laser ablation and fragmentation, their mechanisms, and the inherent advantages of using ultrashort fs laser pulses.

The laser ablation process involves exposing a bulk target to focused fs laser radiation, which scans the surface of the crystalline target immersed in liquid (Figure 2a). This interaction causes material to eject from the target into the liquid, undergoing various transformations before eventually forming colloidal NPs [26,38]. Using fs pulses is particularly advantageous in this method, as it minimizes thermal effects, thereby preserving the structural integrity of the ablated material [30,42]. For instance, a recent study has demonstrated that fs laser ablation can produce spherical TMDC NPs, which retain the original crystal structure and display resonant optical properties [45].

When the fs laser pulse ($>10^9$ W/cm$^2$) interacts with the target surface, it generates plasma originating from the initial material with extremely high local temperature and pressure. As the plasma decays, it transfers energy to the liquid, forming a cavitation bubble [53]. The vapor inside the bubble encloses a region in which solid crystallization and formation of atom clusters occur (Figure 2a). The combination of high pressure, temperature, and plasma density, followed by rapid cooling, creates an ideal thermodynamic environment for the formation of metastable phase nanomaterials (e.g., hexagonal close-packed Ni, Au-Si alloy) [54,55], which are challenging to achieve through conventional chemical methods. During laser ablation in liquids, a small fraction of the ablated material undergoes chemical oxidation at the NPs' surface, even for noble metals [42]. These surface defects are crucial for imparting surface charge and ensuring electrostatic stabilization of the colloids. Nonetheless, most of the colloidal material retains the chemical composition of the initial targets, including multi-component vdW materials.

Alternatively, laser fragmentation involves directing laser pulses into a liquid containing a homogeneously distributed vdW material in powder form [56]. In this process, NP formation is triggered by microparticles within the liquid absorbing the laser energy. Fs pulses induce nonlinear phenomena in the liquid, including self-focusing, filamentation, and the generation of a white light supercontinuum, which enhance energy absorption and promote NP formation. This broad wavelength range allows the initial radiation energy to be effectively absorbed by nearly any source material (Figure 2a). The primary mechanisms of NP formation in this process are Coulomb explosion and photothermal evaporation. According to the liquid-drop model for the Coulomb explosion, a multiply charged particle ejected from the initial educt becomes unstable when the disruptive Coulombic force of charge repulsion exceeds the attractive cohesive force. While this mechanism is dominant at high laser intensities typical for the fs regime and relatively small particle size ($< 50$ nm), the photothermal evaporation is mainly related to the ns pulse duration and large educt particles. Therefore, this approach offers a unique opportunity to fabricate NPs in mobile colloidal form with controlled size, uncontaminated surface, and high stability. Additionally, the integration of laser ablation with laser-induced forward transfer technique presents a promising approach for the fabrication of colloidal solutions of monodisperse NPs [57].

We synthesized stable colloidal solutions of NPs from vdW materials using pulsed laser ablation and fragmentation in liquids (*Methods*), applied to solid targets, powders, and solutions (Figure 2a). The characteristics of the NPs—such as size, shape, and specific properties—can be finely tuned by adjusting various parameters. These include the nature of the target material, laser power, exposure duration, and the liquid medium's composition. By tailoring these parameters, NPs can be customized to meet the specific requirements of applications ranging from electronics to pharmaceuticals.

To study the mechanism of $MoS_2$ NPs crystallization in detail and to characterize the influence of the cooling regime on its structure, we utilized classical molecular dynamics (MD) with the state-of-the-art machine-learning interatomic moment tensor potentials (MTPs) [58,59]. This method enables large-scale MD simulations over tens of nanoseconds with accuracy approaching that of density functional theory (DFT). Further details on the calculation parameters are provided in the *Methods* section.

We performed a series of MD simulations cooling the $MoS_2$ melt from T=2300 K down to complete crystallization (Figure S3). The cooling rate was varied in the range of 0.08-2 K/ns (see the video of the MD simulation in the Supplementary Information (SI)). Our results indicate that crystallization can be divided into two competing processes: the formation of ordered $MoS_2$ layers on the droplet surface, aligned parallel to the phase boundary (heterogeneous nucleation – "shell" in Figure 2b), and the formation of randomly oriented crystallites within the bulk (homogeneous nucleation – "inner crystals" in Figure 2b). The interplay of these processes leads to a core-shell structure, where the core-to-shell ratio and the average size of crystallites in the core depend on the cooling rate (Figure S4). Lower cooling rates result in a thicker shell and larger inner crystallites, whereas higher rates produce a thinner shell and smaller inner crystallites. This demonstrates that the cooling process can modulate the nanostructure parameters of NPs and their optical properties.

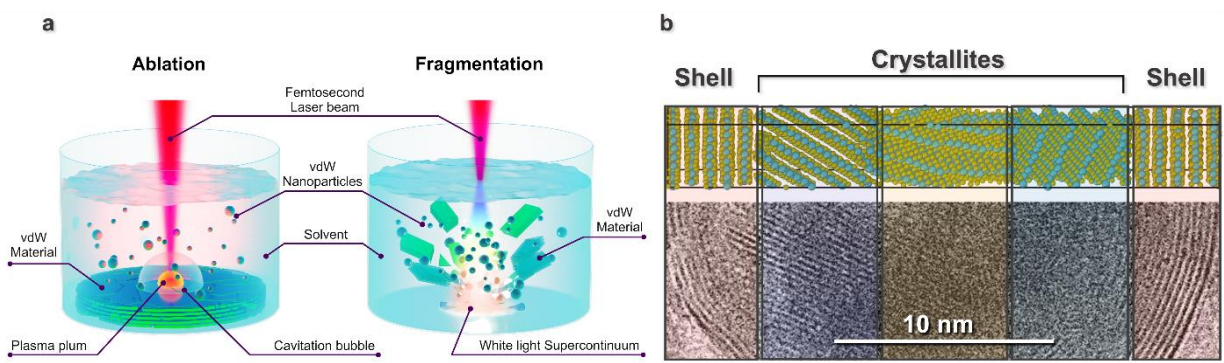

**Figure 2.** (a) Schematic representation of laser-driven synthesis of vdW NPs, (b) atomic structure of core-shell $MoS_2$ polycrystal generated in MD (top) and TEM image of $MoS_2$ NP (bottom).

*The versatility of the laser ablation method for the synthesis of vdW NPs*
Despite the extensive quantitative and structural diversity of vdW materials, laser synthesis in liquid facilitates the production of stable colloidal NPs from a wide range of materials. The laser-matter interaction mechanism, which involves material removal from the target surface, highlights the versatility of this approach. Moreover, the flexibility of laser synthesis in liquid enables processing of metals, semiconductors, and dielectrics in different forms, such as bulk crystalline targets, powders, and liquid suspensions.

In this study, we present the synthesis of geometrically precise NPs from over 50 materials. These materials include TMDCs (sulfides, selenides, tellurides, and Janus structures), M(A)Xenes, and other promising compounds such as hBN, $MoO_3$, $MAPbBr_3$, and CrSBr. Figure 3 depicts the most prominent vdW materials along with their crystal structures, photos of the initial materials, and TEM images of the resulting NPs, thereby confirming the versatility of the laser synthesis method. A comprehensive overview of all synthesized materials, categorized by class, is provided in Figure S1.

In addition to the fundamental ability to synthesize colloidal solutions of vdW NPs, we demonstrate the capacity to control the size and optical properties of the resulting nanomaterials. By varying the liquid medium, we can tune the size distribution, compositional properties, and optical extinction spectra of the NPs. For instance, employing laser ablation of $MoS_2$ in different solvents yields NPs of various sizes: (75 ± 41 nm) in water, (42 ± 24 nm) in acetone, and (15 ± 9 nm) in acetonitrile (Figure S2a). Furthermore, altering the liquid medium type during synthesis allows modulation of the optical and compositional properties of the NPs. Laser fragmentation of $WS_2$ in acetonitrile produces ultra-small quantum dots exhibiting a distinct excitonic resonance in the near-infrared (NIR) region, peaking at around 670 nm (Figure S2b). The choice of solvent significantly affects the extinction characteristics in this spectral range. Specifically, synthesis in acetone produces $WS_2$ NPs with a peak in extinction signal near 750 nm, whereas water-based fabrication results in an extinction spectrum resembling Rayleigh scattering. Additionally, the type of solvent influences the oxidation or carbonization states of the NPs (Figure S2c). Notably, laser synthesis offers the advantage of producing stable NPs with uncompensated surface charges, which was validated across various obtained vdW NPs through Zeta-potential measurements (Figure S2d). This research expands the utility of laser synthesis in advancing nanomaterial development, offering a spectrum of properties and potential applications across diverse fields.

*Morphology and structure of vdW NPs*

During the experiments, we observed that within seconds of operating the high-repetition-rate laser, the liquid medium rapidly changed color, indicating the formation of NPs. The colloidal solution color varied by material and liquid medium, from greenish-brown for $MoSe_2$ in acetone to sky-blue for $Ti_3C_2$ in water. These distinct colors suggest that the NPs retain the optical characteristics of their source materials. We recently demonstrated that a simple centrifugation step can effectively isolate different size fractions of $MoS_2$ NPs [45]. At the same time, variation in the duration of the $MoS_2$ powder fragmentation opens avenues for extra control of the dimensional, optical, and compositional characteristics of the resulting NPs [60]. These modalities unlock the ability to produce 5-10 nm quantum dots with pronounced exciton resonances in the NIR region and spherical NPs exhibiting Mie resonances in the visible and NIR regions [45,60].

To thoroughly investigate the morphology and structure of the laser-synthesized vdW NPs, we utilized high-resolution transmission electron microscopy (TEM), high-angle annular dark-field imaging (HAADF), selected area electron diffraction (SAED), and energy-dispersive X-ray spectroscopy (EDX) analysis (*Methods*). The results indicate that NPs synthesized from various vdW materials—including TMDCs, M(A)Xenes, perovskites, transition metal oxides, and Janus structures—exhibit high crystallinity, retaining crystallographic planes and chemical compositions corresponding to the initial materials (Figure 4; Table S1).

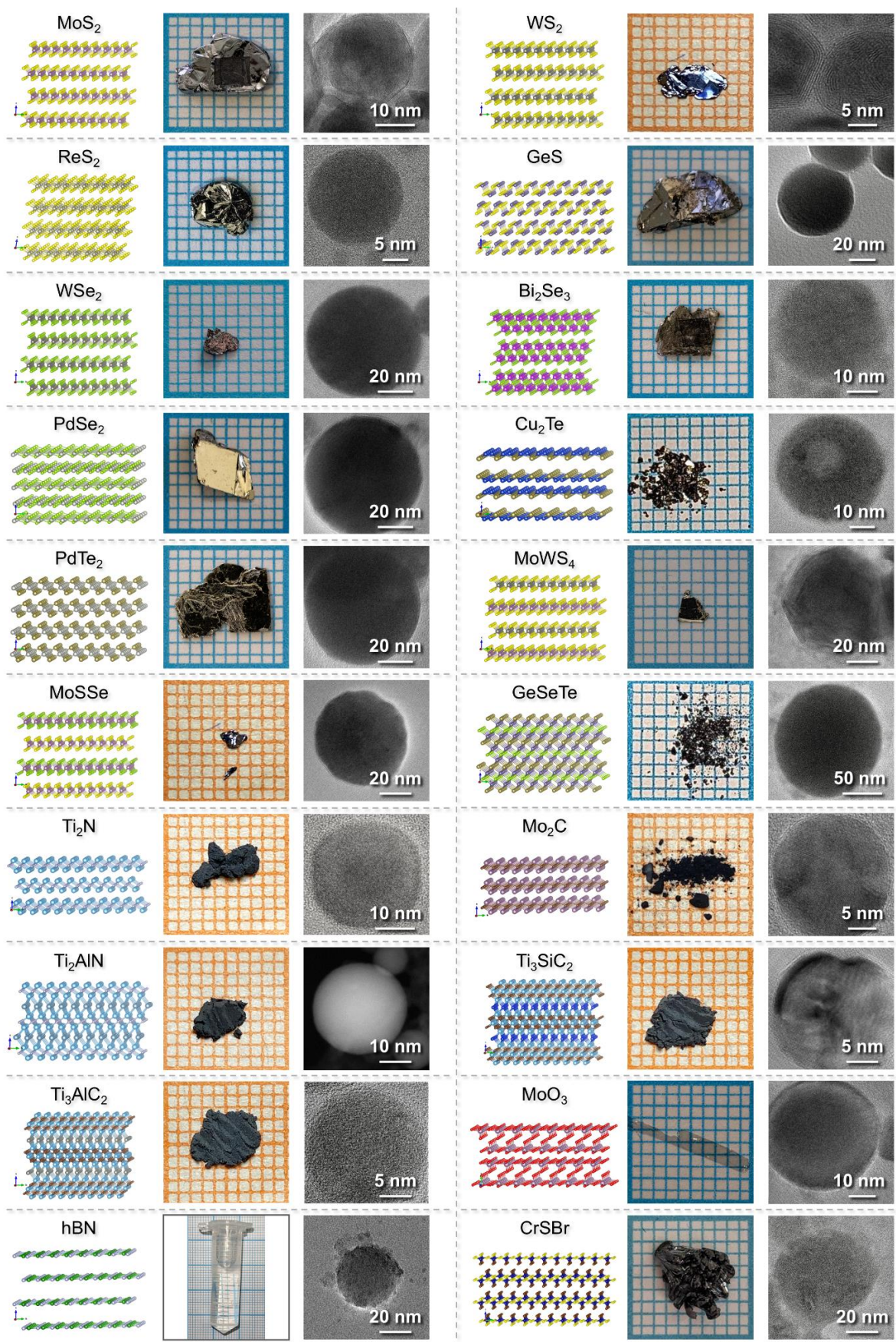

**Figure 3.** Three-dimensional representation of the crystal structure of vdW materials, pictures of initial crystals/powders/solutions, and TEM image of laser-synthesized vdW NPs.

Notably, nanostructures from different materials displayed diverse morphologies, ranging from spherical and fullerene-like shapes to nanotetrahedrons, cubes, and polygons (Figure 4). For instance, laser-ablated $MoSe_2$ (Figure 4a) and $PtTe_2$ (Figure 4c) NPs exhibited fullerene-like polygonal structures that were nearly spherical. TEM images revealed the layered crystalline structure of these nanostructures. Although EDX analysis and SAED patterns confirmed chemical composition consistency, HAADF imaging showed that $MoSe_2$ and $PtTe_2$ NPs had unique interplanar distances in their crystalline shells, which differ from those of the initial materials. This phenomenon appears to be a distinctive feature of the laser synthesis process. Conversely, laser fragmentation of $Ti_3C_2$ powder resulted in the formation of purely spherical single-crystalline NPs with an interplanar distance between layers at the surface of 2.6 Å (Figure 4e). EDX analysis verified that the atomic composition matched the original material. Additionally, SAED analysis confirmed the high crystallinity of the NPs and the correspondence of the diffraction rings to the pure $Ti_3C_2$ diffraction pattern (Table S1).

A different situation was observed for $HfS_3$, $WTe_2$, and perovskite $MAPbBr_3$ (Figure 4b,d,f). In the case of $HfS_3$, we obtained polygonal nanostructures with 6 to 12 facets. A significant quantity of spherical NPs, which is the most typical morphology for laser synthesis in liquid, was also present. These NPs exhibited a pronounced crystalline layered shell with an interplanar distance of 6.4 Å. HAADF analysis and SAED characterization confirmed the high crystallinity of the resulting NPs and the similarity of diffraction patterns to the initial material. EDX images revealed the material composition, consisting of Hf and S.

In contrast, laser ablation of a solid $WTe_2$ crystal resulted in the formation of equilateral nanopyramids with side lengths of 180 nm. These nanopyramids exhibited a stepped structure with a pitch of 7.5 nm, as evident in the TEM image. HAADF analysis revealed the perfect crystallinity of the nanopyramids, while EDX and SAED confirmed the intact chemical composition. Another noteworthy result was the formation of cubic NPs based on perovskite $MAPbBr_3$. These nanostructures also had a pronounced crystalline shell, clearly visible in TEM and HAADF images. Most nanocubes exhibited beveled or rounded corners, likely a feature of the highly nonequilibrium processes of NPs formation during the laser ablation. EDX analysis confirmed the consistency of the original composition, showing the presence of Pb, Br, and N. A detailed analysis of the SAED pattern is presented in Supplementary Table S1.

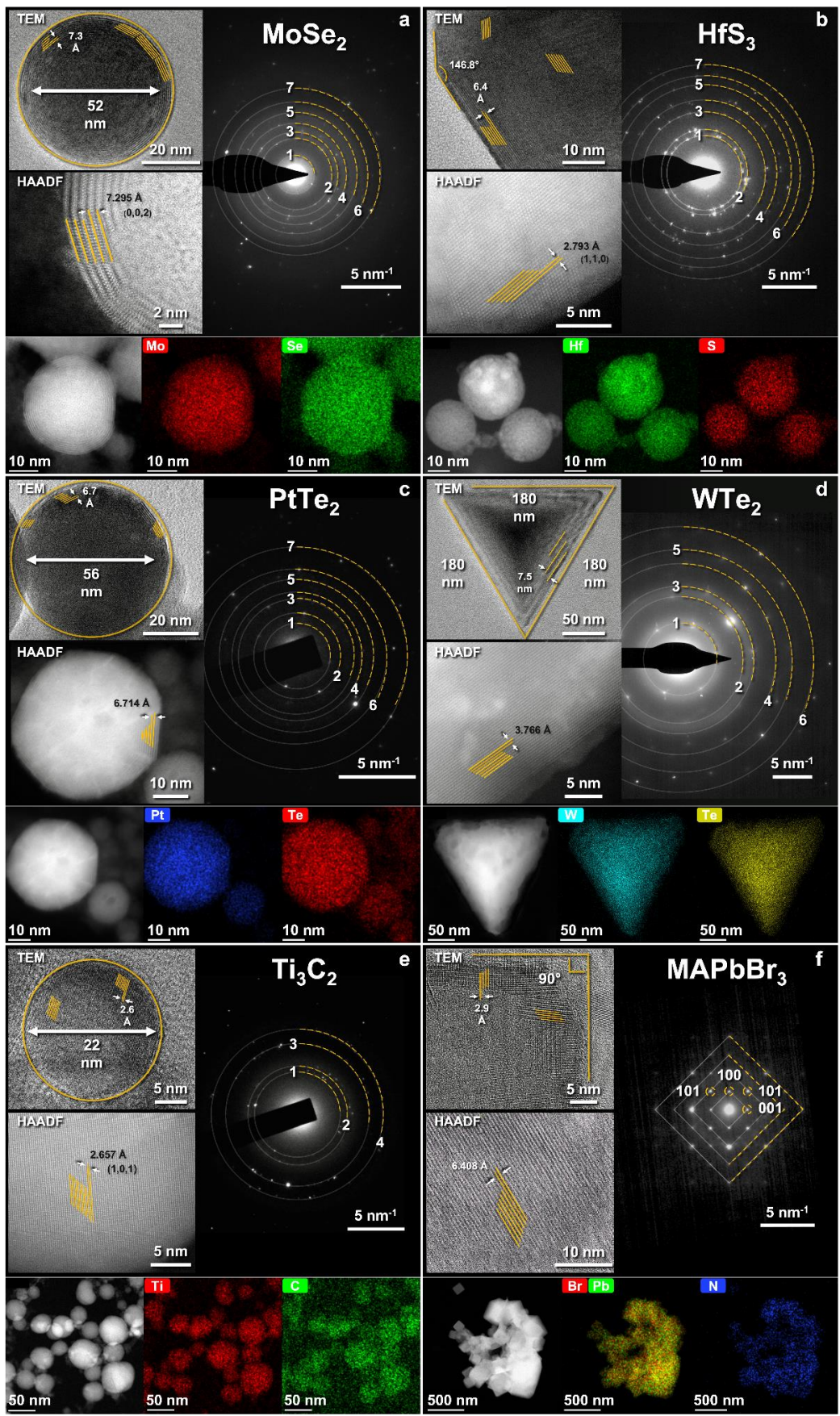

**Figure 4.** Structural and compositional analysis of laser-synthesized vdW NPs. Each panel includes TEM, HAADF, SAED characterizations, and EDX analysis.

Despite the variety of morphologies observed above, spherical NPs remain the most dominant and typical shape of nanomaterials produced by laser synthesis (Figure 3). Moreover, different morphological fractions of laser-synthesized NPs can be easily isolated by centrifugation due to their various size characteristics. For instance, spherical NPs typically exhibit relatively small sizes, controllable within the 2-100 nm range. SEM and TEM analyses revealed that the obtained vdW nanostructures exhibit a lognormal size distribution, with modes generally less than 50 nm and relatively narrow size distributions within the spherical fraction (Figure 5). Specific size distribution characteristics (mode ± FWHM) include (44 ± 18) nm for $MoSe_2$ (Figure 5a), (19 ± 11) nm for $PtTe_2$ (Figure 5c), (50 ± 20) nm for $Ti_3C_2$ (Figure 5e). NPs based on $HfS_3$, $WTe_2$, and $MAPbBr_3$ demonstrated a bimodal size distribution with values of (32 ± 18) nm and (110 ± 55) nm (Figure 5b), (17 ± 6) nm and (144 ± 86) nm (Figure 5d), (13 ± 9) nm and (120 ± 74) nm (Figure 5f), respectively. The second peak is associated with the presence of polygonal nanostructures of $HfS_3$ (Figure 5b, inset 2), nanopyramids of $WTe_2$ (Figure 5d, inset 2), and nanocubes of $MAPbBr_3$ (Figure 5f, inset 2), reflecting a different morphological mode typical for a particular material.

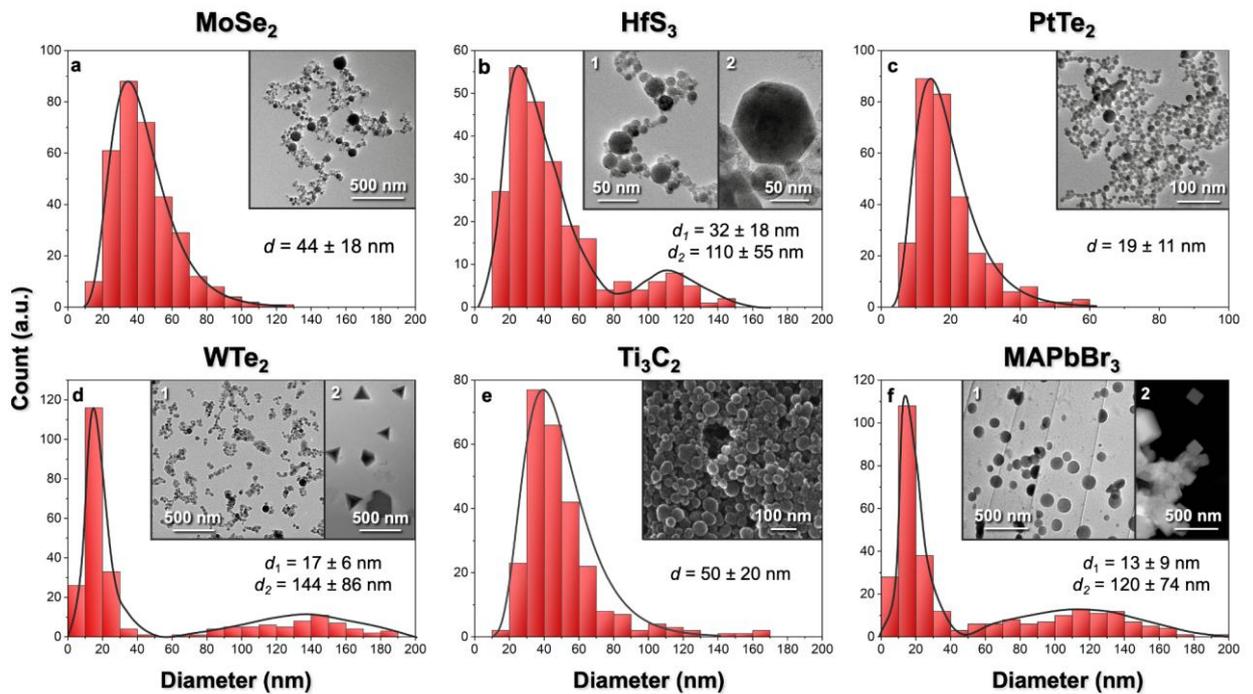

**Figure 5.** Size distribution and representative TEM/SEM image of laser-synthesized vdW NPs.

Thus, we have demonstrated that a relatively simple method, specifically laser ablation or fragmentation, can be effectively used to synthesize NPs from a wide range of vdW materials. A distinguishing feature of these NPs is their well-defined geometric shapes, controllable sizes, and resistance to aggregation. Combined with the inherent unique properties of vdW materials, these characteristics unlock vast potential for their application across various fields. In the following section, we will briefly explore some of the most promising areas for their use.

## Discussion

The potential to fabricate NPs with well-defined shapes from vdW bulk materials introduces a plethora of opportunities across various technological fields due to their distinctive properties, such as semiconductor behavior, extensive surface area, and tunable band gaps. By manipulating the size, shape, and composition of these NPs, their properties can be finely tuned to meet specific needs, making them highly desirable for advanced applications. These NPs are already compatible with various technologies that utilize precisely shaped NPs [45,61]. Figure 6 illustrates some of the most promising sectors for the application of vdW NPs, including diverse electronics, environmental technologies, novel computing, sensors, biomedicine, and more. Below, we will briefly explore some of these NPs' most intriguing application possibilities.

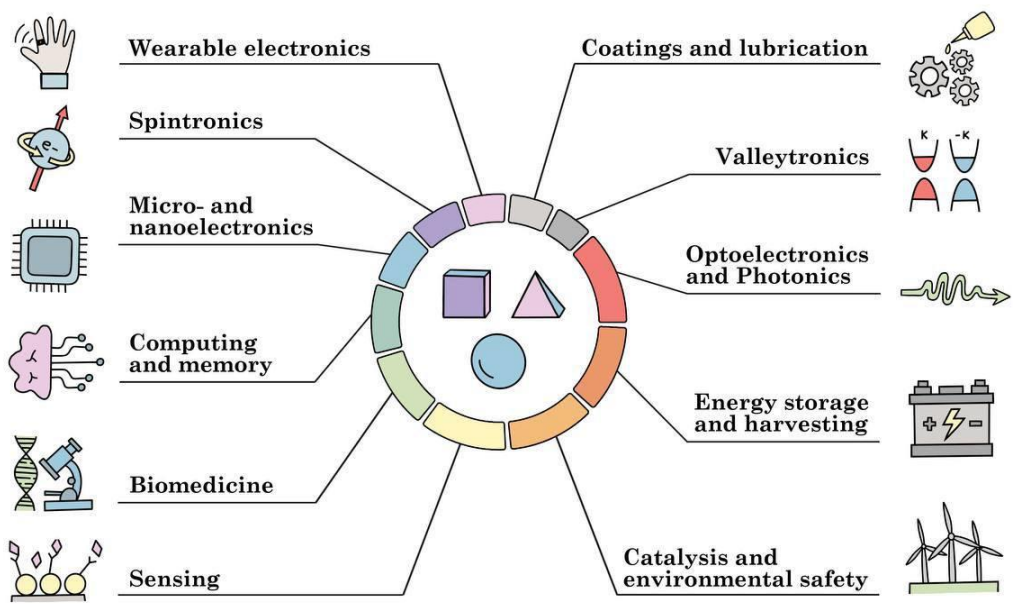

**Figure 6.** Potential applications of geometrically precise vdW NPs. The pie chart categorizes the application sectors based on an analysis of over 3,000 scholarly articles related to TMDC nanoflakes or nanomaterials. The relative sizes of the pie chart segments qualitatively represent the frequency of each application mentioned in the literature. The "Other" category is not displayed in this diagram.

**Catalysis.** Photocatalysis is a prominent method for addressing water pollution through natural sunlight [62]. vdW NPs exhibit exceptional catalytic properties due to their high surface area and active sites [45]. By adjusting the shape and composition of these NPs, their catalytic activity and selectivity for specific reactions can be optimized, potentially leading to more efficient and sustainable industrial processes. Moreover, vdW materials are gaining recognition for their ability to generate hydrogen via photocatalytic water splitting, presenting a more economically viable alternative to traditional photocatalysts [63,64]. For example, $MoS_2$ and $MoSe_2$ are notable for their widespread availability, affordability, robust physicochemical stability, and low bandgap energy (1.6 eV), making them prime candidates for such applications [65,66].

Additionally, vdW materials hold promise for water electrolysis. For instance, due to a lower onset potential, a combination of $MoS_2$ NPs with graphene-like substances [67] enhances the HER activity beyond that of standalone $MoS_2$. The smaller size of TMDC NPs contributes to their extensive surface area, reducing the distance for charge transfer and thereby accelerating the water-splitting process. A large specific surface area favors the adsorption of organic pollutants and facilitates electron transport [68]. Consequently, developing an economical method for producing such NPs is crucial for both scientific and practical reasons. Laser ablation has emerged as a highly effective technique in this context, underscoring its significant application value.

**Sensing.** In environmental monitoring, food quality assurance, healthcare, pharmaceutical analysis, and other fields, the role of chemical sensors and biosensors is increasingly critical [69]. Semiconductor-based sensors are appealing due to their cost-effectiveness and simple operational principles, which rely on variations in conductance upon exposure to specific gases [70]. vdW NPs are particularly noted for their applicability across a spectrum of sensing technologies [71]. Numerous vdW-derived NPs have been engineered to refine electrochemical biosensors, given their minimal noise interference and enhanced carrier mobility [72]. Furthermore, vdW materials are excellent for creating energy-efficient, flexible gas sensors that operate at room temperature, a particularly challenging capability with metal oxides.

Studies on 2D TMDCs such as $MoS_2$, $MoSe_2$, $WS_2$, and $WSe_2$ have shown that these materials surpass others in sensitivity, selectivity, and response speed, making them ideal for flexible and wearable gas sensors [73]. When NPs are combined with materials like graphene, gas detection sensitivity can be further enhanced. Additionally, vdW NPs can be used for electrochemical biosensors for DNA analysis when incorporated into composites with graphene or CNTs [74]. Photoelectrochemical sensing, which operates under visible light, requires photocatalysts such as vdW NPs. Notable innovations include $WS_2$-graphene nanosheets and gold NPs for the label-free electrochemical determination of IgE [75]. Moreover, $MoSe_2$-graphene composites have shown efficacy in sensing PDGF-BB [76]. Consequently, vdW NPs exhibit significant potential for advancing a broad array of sensing applications, underscoring their critical role in the future development of sensor technology.

**Energy.** In the energy sector, vdW NPs play a crucial role in enhancing the performance of energy storage and conversion devices, such as lithium-ion batteries (LIBs), supercapacitors, and solar cells. The large surface area of the NPs improves interaction with electrolytes, boosting both the storage capacity and rate capability. Additionally, the tunable band gaps of TMDCs allow for the fine-tuning of solar cells' absorption spectra, potentially elevating their efficiency [77]. For instance, materials such as $MoS_2$ and $WS_2$ possess these properties and a high density of reactive sites at their edges, making them ideal for electrochemical reactions. The interlayer spacing in TMDCs accommodates lithium ions without significant volume expansion, outperforming silicon-based alloys. Studies have shown [78] that $MoS_2$, combined with carbon, has enhanced performance, underscoring its reliability as a material for flexible LIBs [79]. The exceptional surface-to-volume ratio and uniform shape of vdW NPs are expected to improve LIB performance further, providing superior control and reproducibility.

For supercapacitors, where the electrode's composition is critical to the system's overall performance [80], vdW materials are also highly effective. Although carbon-based materials are commonly used as electrodes, they suffer from low energy density. TMDCs, including $WS_2$ and $MoS_2$, provide numerous active sites for charge storage and channels for ion exchange, leading to significantly high specific capacitance due to rapid redox reactions. The electrode material's high surface area and porosity enhance ion transport, thereby increasing capacitance [81]. The electrochemical properties of geometrically precise vdW NPs, especially when combined with carbon materials, are considered superior, although extensive research is needed to validate this.

**Computing.** vdW NPs, with their adjustable band gaps, are pivotal in advancing next-generation electronics and optoelectronic devices. This is especially true in the emerging field of bio-inspired neuromorphic computing systems (NCSs) and resistive random-access memory (RRAM) devices [82]. vdW materials bridge the gap between conventional and neuromorphic electronics as foundational components for memristors, thin-film transistors, photodetectors, and other essential devices. NCSs mimic human neural architectures through computing-in-memory [82]. Memristors, in turn, emulate synapses via resistive switching (RS) effects, promoting a shift from von Neumann's paradigm [83]. This transition supports the development of devices for diverse applications in robotics, the Internet of Things (IoT), biomedicine, and more [84]. The high electronic mobility, exceptional light absorption, and tunable conductance of vdW materials make them ideal for RS technology, enabling ultra-low power consumption.

Numerous vdW materials have been proposed as the basis for memristors [85]. Notable examples include $MoSe_2/MoS_2$ heterostructures [85]; $MoSe_2$ nano-islands [86], nanorods [87], nanosheets [88], and nanocomposites [89]; $MoSe_2$/PVA systems [90] with varying NPs [91] concentrations; and guar gum-$WTe_2$ nanoflakes hybrids [92]. Structures based on vdW materials can also be utilized as photonic electrical switches [93]. For instance, an NIR photonic RRAM device incorporating $MoS_2$ with $NaYF_4$:$Yb^{3+}$, $Er^{3+}$ upconversion NPs, demonstrated stable NIR light-controlled RS [94]. Additionally, $Bi_2S_3$ and $Bi_2Se_3$ NPs, known for their strong NIR absorption, also indicate potential applications in NIR RRAM [95,96]. Utilizing NPs increased the maximum $I_{on}/I_{off}$ ratio [90] and decreased the operating voltages [92] of memristors, thus indicating the superiority of NPs in this area. Using NPs of the correct shape may enhance RS's stability and reproducibility, addressing a significant challenge in contemporary memristor devices. In conclusion, vdW NPs can significantly advance NCSs and RRAM devices, enhancing performance and efficiency in next-generation electronics and optoelectronics.

**Nanocomposites.** Like other NPs, vdW materials possess a broad potential for use as nanocomposites, a potential that is yet to be fully explored due to the vast number of possible combinations. For instance, in photocatalysis, heterojunction composites benefit from the disparate energy levels of two semiconductors merging to create a system that promotes swift and effective charge separation, reducing the recombination rate of electron-hole pairs [65]. This leads to a notable enhancement of photocatalytic activity. Similar charge separation strategies have been applied to traditional oxide composites like $TiO_2/MoO_3$ [97]. Various TMDC composites have been evaluated for their photocatalytic efficiency under visible light, such as $MoS_2$/BiOBr [98], $MoS_2$/rGO [99], and $MoS_2/C_3N_4$ [100]. Nanocomposites, where NPs are embedded into a matrix, such as a polymer, have also demonstrated their potential in the development of energy-efficient memristors, as seen in the $MoSe_2$/PVA composite [90]. Huge opportunities also arise from combining TMDCs, MXenes, and perovskites, especially with polymers, for various biomedical applications, sensing, industrial uses, and more. Controlling the shape and size of the particles will enable the achievement of stable and reproducible results.

**Biomedicine.** The tunable optical and chemical properties of vdW-based nanostructures extend their biomedical applications beyond sensing to include advanced therapies such as drug delivery, photothermal therapy (PTT), photodynamic therapy (PDT), and radiation therapy (RT), as well as diagnostics like computed tomography (CT), magnetic resonance imaging (MRI), and photoacoustic imaging (PAI) [101,102]. Unlike traditional plasmonic metals, vdW nanomaterials absorb dominantly in the NIR range, enhancing PTT and PAI for non-invasive theranostics of deep-seated tumors [101–104]. For instance, we recently demonstrated that $MoS_2$ NPs have shown nearly double the photothermal response compared to Si NPs [45].

Pronounced NIR absorption and high catalytic activity of vdW materials significantly enhance the efficiency of PDT [105,106]. Additionally, vdW nanomaterials with high atomic numbers (e.g., $WS_2$, $Ta_2C$) serve as CT contrast agents and RT sensitizers, while those with magnetic properties (e.g., $Cu_{2-x}S$) function as MRI contrast agents [107,108]. The multimodality and synergy with conventional therapies, combined with high biocompatibility and the potential for biodegradation of vdW NPs, appeal as a very promising basis for the medicine of the future [109,110]. The ability to engineer their small size and surface chemistry for optimal biodistribution, target specificity, and drug release profiles highlights their potential to revolutionize advanced personalized medicine.

**Additive technologies.** Additive manufacturing techniques have revolutionized the production of composite materials, offering a straightforward solution for creating complex, custom-designed shapes. In the biomedical sector, 3D printing has enabled the fabrication of scaffolds tailored to patient-specific needs, allowing for precise control over product architecture and microstructure [111]. Polymers reinforced with graphene or its derivatives have shown significant potential for applications requiring enhanced electrical and mechanical properties and improved cell response, thereby becoming increasingly attractive for biomedical applications [112]. The ability to tailor the properties of vdW NPs through controlled synthesis and functionalization opens new avenues for innovation in material design and manufacturing. As research in this field progresses, the integration of vdW NPs into additive technologies is anticipated to drive significant advancements in the fabrication of next-generation devices and systems. Laser-induced forward transfer (LIFT), a technique that involves the use of focused laser pulses to transfer material from a donor to a receiver substrate in a controllable manner, may benefit from using the vdW materials. Recent development in the LIFT technique has led to the controlled fabrication of metallic (Au, Ag, Al, Cu, Fe, etc.) and semiconductor (Si, Ge, etc.) NPs with precisely adjustable radii between 50 nm and 1 um and their accurate positioning on the desired substrate [57]. The use of vdW films as donor substrates for printing vdW NPs could significantly expand the applications scope of nanophotonic structures produced by LIFT owing to the high refractive index, low ohmic losses, and giant birefringence of many vdW materials [113]. Beyond laser-based techniques, vdW NPs have potential in other additive manufacturing methods. For example, in inkjet printing, the dispersibility and stability of vdW NPs in various solvents can lead to the development of high-performance inks for printing high-refractive index coatings and sensors.

**Thermoelectricity.** vdW materials are known for their superior thermoelectric (TE) characteristics, attributed to their low dimensionality in electronic and phonon transport. Studies have demonstrated significant enhancements in the TE properties of 2D $WSe_2$ and $WS_2$ compared to their bulk counterparts [114]. Other promising materials include $PtSe_2$ [115] and $PdSe_2$ [116]. It has been suggested that due to a variety of phonon scattering mechanisms and the inherent energy dependency of their electronic density of states, low-dimensional materials may exhibit improved TE performance over bulk materials [117]. The TE efficiency of $MoS_2$ has also been shown to surpass that of 3D semiconductors [118]. Furthermore, $Bi_2S_3$ NPs have been explored for their TE properties, showing promise for use in waste heat recovery and cooling applications [119]. The combination of low thermal conductivity and tunable electrical properties positions $Bi_2S_3$ as a strong candidate for efficient TE materials capable of transforming temperature gradients into electrical energy [120].

**NIR shielding.** The development of efficient NIR shielding materials is crucial for reducing heat absorption and providing protection against NIR radiation, offering substantial benefits in terms of energy conservation and personal safety [121]. The pronounced absorption observed in various vdW materials, such as $Bi_2S_3$, within the NIR spectrum renders them highly suitable for NIR shielding applications. This characteristic is particularly valuable for the development of protective coatings for buildings or vehicles aimed at diminishing heat intake from sunlight. Such advancements contribute to significant energy savings in air conditioning systems. Integrating an infrared shielding function with photocatalytic materials paves the way for multifunctional materials, envisaged as innovative smart window coatings. Considering the exceptional photocatalytic properties of vdW NPs, their application in NIR shielding coatings emerges as a promising avenue.

In summary, our approach to producing geometrically precise vdW NPs holds significant promise across a broad range of technological applications due to their unique and tunable properties. Their development and integration into devices and systems have the potential to drive innovations and improve performance in energy, catalysis, medicine, neuromorphic electronics, environmental remediation, and beyond. However, despite the promising prospects, several challenges need to be addressed. These include the scalability of the synthesis process of NPs with controlled size, shape, and composition, as well as ensuring their stability and compatibility in various applications. Another drawback of many powder materials, including NPs, is their challenging recyclability and reusability. This limitation not only contributes to secondary environmental pollution but also incurs significant costs due to the need for complex separation technologies. Continued research and development are crucial to address these challenges and fully exploit the potential of vdW NPs.

## Conclusions

The emergence of advanced technologies for the fabrication of layered vdW materials with diverse compositions and tunable physicochemical properties has ushered in a new era for both scientific research and industrial applications. Due to their synthetic nature, these materials present unprecedented opportunities for developing over 5,000 nanostructures that achieve remarkable performance in fields such as energy applications, catalysis, biomedicine, additive manufacturing, nanophotonics, computing, and sensing. Despite these advances, a significant challenge remains in the miniaturization of technologies and functional materials, particularly the lack of a universal method for controllably modifying the dimensional, morphological, and optical properties of vdW nanomaterials while preserving their original composition and stability in dispersed systems.

In this study, we address this challenge by demonstrating, for the first time, the remarkable versatility of the fs laser synthesis method for the production of colloidal NPs from vdW materials and perovskites. Our approach allows for precise control over the morphology, size, composition, and optical properties of NPs, achieving high colloidal stability and maintaining the original crystalline structure of over 50 vdW materials. We have shown that laser synthesis can produce a diverse array of nanostructures, including fullerene-like, polygonal, and pyramidal shapes from TMDCs, single-crystalline NPs from M(A)Xenes, and crystalline nanocubes from perovskites. Molecular dynamics simulations further elucidate the crystallization processes involved, revealing that both heterogeneous and homogeneous nucleation mechanisms contribute to the formation of these nanostructures, with the size and core-shell ratio being dependent on the cooling rate.

The successful demonstration of this method not only provides a new tool for the fabrication of vdW-based NPs but also opens vast potential for further innovations in scientific research and technological applications. The ability to create such diverse and well-defined nanostructures paves the way for advancements in numerous fields and offers exciting prospects for future exploration beyond the current applications of layered materials.

## Author Contributions

G.I.T., A.V.K., B.N.C., A.V.A., K.N., and V.S.V. suggested and directed the project. A.A.M., G.A.E., G.V.T., I.S.K., D.V.D., D.A.P., D.I.T., A.A.P., and A.V.S. performed the measurements and analyzed the data. A.B.M., S.S., N.D.O., and I.A.K., provided theoretical support. Z.S., F.L. and U.A. synthesized part of vdW crystals. G.I.T., A.A.M, G.A.E., G.V.T., I.S.K., and D.A.P. wrote the original manuscript. All authors reviewed and edited the paper. All authors contributed to the discussions and commented on the paper.

## Competing Interests

The authors declare no competing interests.

## Acknowledgments

Z.S. was supported by ERC-CZ program (project LL2101) from Ministry of Education Youth and Sports (MEYS) and by the project Advanced Functional Nanorobots (reg. No. CZ.02.1.01/0.0/0.0/15_003/0000444 financed by the EFRR).

The authors thank Dr. Valentin Solovey, Elizaveta Gordeeva, and Daria Oliinik for their help in creating the illustrations.

## Data Availability

The datasets generated and analyzed during the current study are available from the corresponding author upon reasonable request.

## Methods

*Laser-assisted synthesis of vdW NPs*

The colloidal solution of 2D material NPs was prepared using the femtosecond (fs) laser fragmentation and ablation in liquids, similar to our previous studies [45,60,122,123]. A 3 mm beam from Yb:KGW system (1030 nm, 250 fs, up to 400 µJ, 1-200 kHz, TETA-20 model, Avesta, Russia) was used as a laser radiation source. For powdered-type materials, the laser fragmentation technique was used. Here, the reshaping and size reduction processes are based on the interaction of so-called white light supercontinuum irradiation with material (Figure 2a). In case of fragmentation, the initial powder was dispersed in 10-20 ml of acetonitrile (purity ≥99.9%, HPLC Grade, J.T. Baker, USA) by ultrasonification (5 min) in a glass chamber (BK-7, wall thickness 3 mm). Acetonitrile was used to preserve the original material composition and reduce possible oxidation of the NPs surface, typical for laser synthesis in liquids with a high oxygen content [42,124]. The resulting solution was then irradiated with fs laser pulses, and the focus was adjusted so that the supercontinuum could be generated at 1 cm from the entrance wall of the cuvette. The laser pulses' energy was 50-100 μJ, and the repetition rate was 1-100 kHz. Homogenization of the solution during the fragmentation process was done by magnetic stirrer and manual mixing with a pipette. The concentration of the initial colloidal solution before fragmentation was 0.1 g/L. To increase fragmentation efficiency and utilize most of the volume, we continuously moved the laser beam over the solution using a galvanometric scanner (LScan-10, Ateko TM, Russia) at a speed of 4 m/s. The duration of the laser fragmentation step ranged from 5 to 30 minutes. The relatively large NPs (>200 nm) were sedimented by a centrifugation step (7,500 g, 1 min, liquid height 50 mm, 24 °C) and then fragmented again.

In the case of bulk crystalline materials, the technique of laser ablation in liquids was used. The initial target was fixed vertically in a PTFE stand inside a glass chamber filled with 10-50 ml of acetonitrile. The laser beam was focused on the target surface by the F-Theta lens (63 mm focal distance, Thorlabs, USA). The liquid thickness between the target and chamber wall was minimized to 3 mm to improve the synthesis productivity. The pulse energy was 5-10 μJ, and the repetition rate was 1-200 kHz. To avoid ablation from one spot and increase efficiency, we continuously moved the laser beam over the target surface using the galvanometric scanner at a speed of 5 m/s. The duration of laser ablation was in the range of 5-30 minutes. The relatively large NPs (>200 nm) were sedimented and removed by a centrifugation step (7,500 g, 1 min, liquid height 50 mm, 24 °C).

*Characterization of vdW NPs*

The size characteristics and atomic composition of the synthesized NPs were studied utilizing a scanning electron microscopy (SEM) system (MAIA 3, Tescan, Czech Republic) operating at 30 kV coupled with an EDX detector (X-act, Oxford Instruments, UK). Samples for SEM imaging were prepared by dropping 1-5 μL of the NPs solution onto a cleaned silicon wafer with subsequent drying at ambient conditions.

Morphological and structural properties of synthesized NPs were characterized by the high-resolution TEM system (JEOL JEM 2100, Japan) operating at 200 kV with a GatanMultiscan charge-coupled device in imaging and diffraction modes. Detailed study of the fine structure of the vdW NPs was carried out with the S/TEM (HRTEM, HRSTEM, HAADF) system (Titan Themis Z, ThermoFisherScientific, Netherlands) operating at 200 kV, coupled with EDX detector (Super-X, ThermoFisherScientific, Netherlands). The Titan Themis Z microscope can correct spherical aberrations, which significantly improves the microscope's resolution. Samples were prepared by dropping 2 μL of NPs solution onto a carbon-coated TEM copper grid and subsequent drying at ambient conditions.

The size distribution of the synthesized NPs was obtained by analysis of the SEM images in the ImageJ software environment with circle fit approximation. The final distribution was based on the measurement of 300-500 NPs diameters.

Hydrodynamic size distribution and Zeta-potential were measured by dynamic light scattering (DLS) and doppler-anemometry technique, respectively, with Malvern Zetasizer (Nano ZS, Malvern Instruments, UK). Mode values ± half-width of the peak of number-weighted size distributions were used as the hydrodynamic diameter. Smoluchowski approximation was used for Zeta-potential calculation.

*Molecular dynamics*

MD calculations were performed using the Large-scale Atomic/Molecular Massively Parallel Simulator (LAMMPS) program package [125] with machine-learning interatomic potential MTP [59]. The simulation box with approximate size 203x27x25 Å contained 6912 atoms (see video of MD simulation in the Supporting Information). Pressure and temperature were controlled via the Nose–Hoover thermostat and barostating [126]. The integration timestep was 0.5 fs.

We used an iterative active learning scheme to sample the training data for the interatomic potential MTP. In each iteration, a few atomic configurations of the Mo-S clusters were randomly generated to initialize the parallel sampling from MD trajectories. The sampled configurations were then aggregated into a single list cleaned from duplicates. The energies and interatomic forces were calculated for each configuration using the DFT approach implemented in the VASP package [127]. The training set of the MTP potential was therefore updated, and the potential was retrained on the updated data to start the new sampling iteration. Finally, the sampling was stopped when no new structures were sampled from the MD trajectories for 10 iterations. The resulting training set consisted of 3250 atomic configurations. The error in predicting the energies on the training data was 16 meV/atom.

For the DFT calculations, we used the PBE exchange-correlation functional [128] with PAW pseudopotentials [129] and D3 dispersion corrections [130]. The cutoff energy for plane waves was 600 eV, and a single $\Gamma$ point was used to sample the first Brillouin zone.